\documentclass[runningheads]{llncs}
\usepackage[T1]{fontenc}
\usepackage{amsfonts,amssymb}
\usepackage{pgfplots}
\usepackage{caption}
\usepackage{subcaption}
\usepackage{graphicx}
\usepackage{multirow}
\usepackage{color}
\usepackage{amsmath}
\usepackage{bm}
\usepackage{adjustbox}

\begin{document}
\title{TruthSR: Trustworthy Sequential Recommender Systems via User-generated Multimodal Content}
\titlerunning{TrustSR}
%
%

\author{Meng Yan\inst{1} \and
Haibin Huang\inst{1} \and
Ying Liu\inst{2}\and
Juan Zhao\inst{3} \and
Xiyue Gao\inst{1}\and
Cai Xu\inst{1}\and
Ziyu Guan\inst{1*}\and
Wei Zhao \inst{1}}
\authorrunning{Meng. Author et al.}
\institute{Xidian University, Xi'an, China \\
\email{\{mengyan@stu.,22031212451@stu.,xygao@,cxu@,zyguan@,ywzhao@mail.\}xidian.edu.cn}
\and
Northwest University, Xi'an, China \\
\email{liuying6@stumail.nwu.edu.cn}
\and
Peng Cheng Laboratory, Shenzhen, China \\
\email{Zhaoj09@pcl.ac.cn}}

%
\maketitle      
\begin{abstract}

Sequential recommender systems explore users' preferences and behavioral patterns from their historically generated data. Recently, researchers aim to improve sequential recommendation by utilizing massive user-generated multi-modal content, such as reviews, images, etc. This content often contains inevitable noise. Some studies attempt to reduce noise interference by suppressing cross-modal inconsistent information. However, they could potentially constrain the capturing of personalized user preferences. In addition, it is almost impossible to entirely eliminate noise in diverse user-generated multi-modal content. To solve these problems, we propose a trustworthy sequential recommendation method via noisy user-generated multi-modal content. Specifically, we explicitly capture the consistency and complementarity of user-generated multi-modal content to mitigate noise interference. We also achieve the modeling of the user's multi-modal sequential preferences. In addition, we design a trustworthy decision mechanism that integrates subjective user perspective and objective item perspective to dynamically evaluate the uncertainty of prediction results. Experimental evaluation on four widely-used datasets demonstrates the superior performance of our model compared to state-of-the-art methods. The code is released at https://github.com/FairyMeng/TrustSR.

\end{abstract}

\section{Introduction}
\begin{figure}[t]
\centering
\includegraphics[width=1\textwidth]{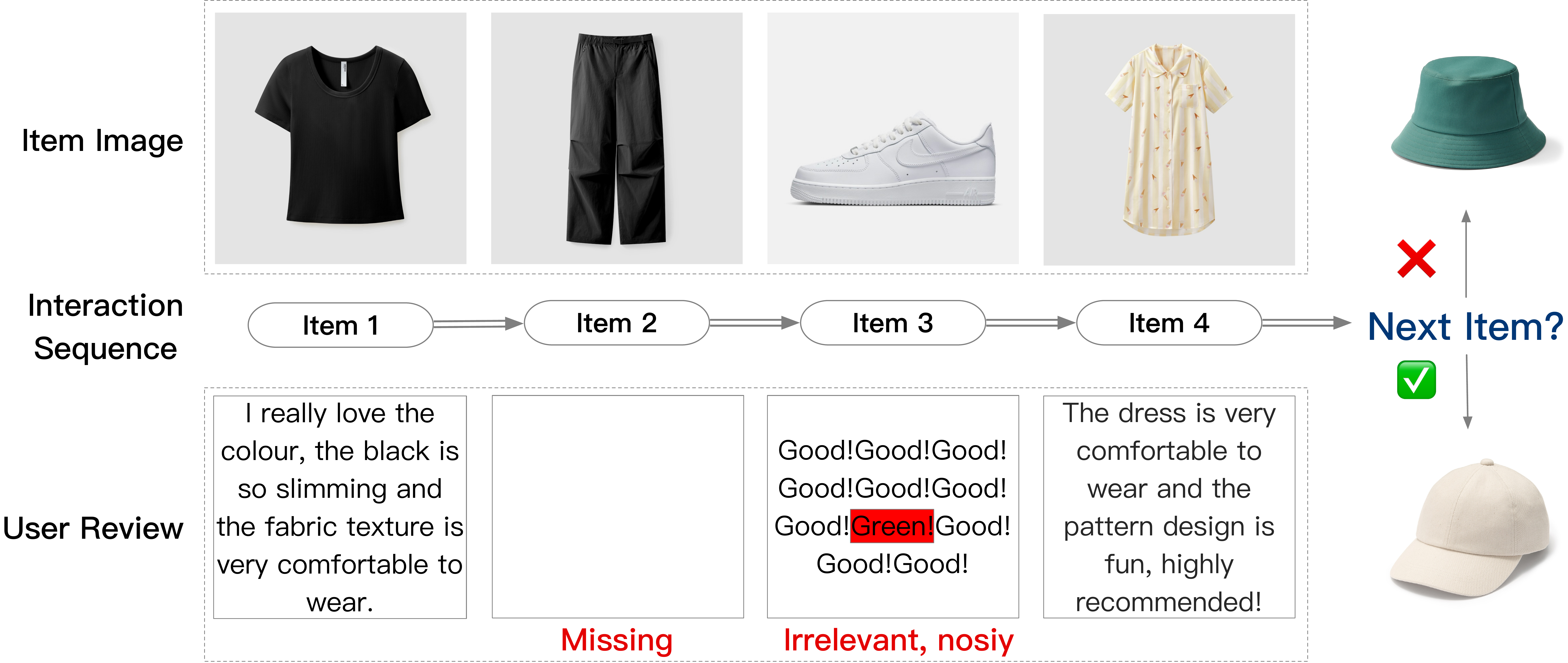}
\caption{Illustration of interference information in UGC. User review on item 2 is \textbf{missing}; user review on item 3 is \textbf{noisy} and \textbf{irrelevant}. The interference information affects the selection of the next item.}
\label{fig1}
\end{figure}

The sequential recommendation takes into account the user historical interactions with items and predicts the next item that the user may find interesting~\cite{ref_introduction7,ref_introduction8}. This approach closely aligns with the user actual behavior patterns and plays a crucial role in enhancing user experience. Achieving accurate predictions in sequential recommendation typically relies on modeling item correlations. Earlier studies~\cite{ref_introduction9} model correlations based on the number of users who have purchased the same items. However, the number of items interacted with by users is insignificant compared to the total number of items, resulting in the problem of sparsity. To address this issue, recent researches~\cite{ref_relate7,ref_relate8,ref_relate13} tackle the sparsity problem by incorporating item side information, such as category, title, home page image, and so on. This additional item side information enriches item representations and improves the modeling of item correlations. Nevertheless, since this item side information is generated by merchants, it is limited in scope and lacks diversity.


In addition to item side information, there are user-generated content (UGC) in recommendation systems. UGC is a form of user-created information that encompasses various types, such as reviews, images, videos, ratings, etc. It serves to supplement fundamental user interactions like purchasing and browsing by providing valuable insights into user preferences and product features. Researchers tend to address the sparsity challenge by extracting substantial and rich information from the plethora of UGC~\cite{ref_introduction2,ref_introduction11,ref_introduction12,ref_introduction13}. Nevertheless, elements such as intrusive advertisements or inaccuracies reviews can introduce a certain level of noise into UGC, potentially impacting the efficacy of recommendation systems. In Fig.~\ref{fig1}, an input error likes "green" in item 3 could lead to the incorrect assumption that the user likes green items.

Some studies delve into the exploration of intrinsically related features across different modalities in UGC, and attempt to reduce noise interference by emphasizing inter-modal consistency information~\cite{ref_UGC+noise,ref_A-semi}. However, it is crucial to acknowledge that there may be differences in characteristics between modalities, and an overemphasis on cross-modal consistency could potentially constrain the capturing of personalized user preference. Moreover, all the above efforts are just trying their best to mitigate the impact of noise, and we should admit that completely eliminating noise from UGC is still an unsolved challenge. An incorrect recommendation result may lead to significant user churn. Therefore, when applying UGC information to sequential recommendations, we should ensure the recommendation accuracy, and more importantly, the corresponding recommendation reliability.


To address the above problems, we propose a Trustworthy Sequential Recommendation method (TruthSR) via noisy user-generated multi-modal content. Specifically, we capture the consistency and complementarity of user-generated multi-modal content to mitigate the noise interference. We also achieve the modeling of the user multi-modal sequential preferences. In addition, we face the problem that noise cannot be completely eliminated, and design a trustworthy decision mechanism that integrates subjective user perspective and objective item perspective to dynamically evaluate the uncertainty of prediction results. Potentially interested items with high confidence can be recommended to users more directly on the first page, while items with low confidence are sorted back to be presented in a more conservative way, ensuring the accuracy and usability of the recommendations. Overall, our contributions are as follows:


\begin{itemize}
    \item {We propose to explicitly mine user multi-modal sequential preferences from their low-quality generated content and show that this idea can significantly improve recommendation performance.}
    \item {We point out that it is almost impossible to entirely eliminate noise in diverse UGC, and the trustworthy decision is essential. To solve this problem, we develop a straightforward yet effective mechanism to ensure decisions are trustworthy.}
    \item {We conduct experiments on four widely used real-world datasets, and experimental results show that TruthSR can significantly improve performance and outperform state-of-the-art methods.}
\end{itemize}

\section{Related Work}
\subsection{Sequential Recommendation} 
Sequential recommendation aims to capture users preferences from their historical sequential interactions and make the next-item prediction. Early sequential recommendation studies~\cite{ref_relate1,ref_relate2,ref_relate3,ref_relate4} are often based on the Markov Chain assumption and Matrix Factorization methods, recommending the next items based on similarity, which always suffers from sparsity. Afterwards, some works attempt to introduce side information to alleviate the sparsity problem.
Zhou et al.~\cite{ref_relate19} model the relationship between item and attribute to enhance data representation. Xie et al.~\cite{ref_baseline5} encode text such as item categories and titles into features, and then use an attention mechanism to enhance the item representation. 
~\cite{ref_relate14} also incorporates visual signals into predictors of people's opinions. 
However, they focus on using item-related side information such as product descriptions, titles, images to enrich item representations, this information is fixed with templates on the item side and lacks diversity. UGC is the user self-created content that covers multiple forms of information, we consider UGC-based sequential recommendation and develop a new module to understand the user preferences and behavioral trends from both user and item view.

\subsection{User-generated content based Recommendation}
Users' textual comments contain rich users preferences, and images contain visual preferences that are difficult to express in language. Recent studies propose many recommender systems based on user-generated content. Huang et al.~\cite{ref_relate17} build a neural network for each modality to extract high-level feature representations, and then fuse these feature representations to obtain multi-modal representations of users and items for recommendation. Zhang et al.~\cite{ref_relate18} capture the association of each image region with each word of the text modality using bi-directional attention. 
Multi-modal data allows richer description of users and objects from multiple dimensions~\cite{ref_AF}.
However, there are two main problems in multi-modal UGC-based recommendation systems: (1) Existing methods generally use UGC directly for model training. However, the large amount of missing, irrelevant, and irregular anomalous data will bring negative effects.
(2) Most of the existing studies only use simple fusion methods, such as high-level feature splicing, when fusing multi-modal data of users/items, ignoring the consistency and complementarity among different modalities. 
In contrast, our approach attempts to capture the consistency of multi-modal content to reduce the influence of noise. We also design a trustworthy decision mechanism that combines two perspectives to dynamically evaluate the credibility of recommendation results, the results are more convincing and less risky when applied to different scenarios.

\section{Problem Description}
Given a user $U$'s historical interaction sequence $S^U=\{ I_1,I_2,...,I_N \} $, we define $R^{U}=\{ r^{U}_1,r^{U}_2,...,r^{U}_N \} $ denotes $U$'s reviews for $S^U$, $R^{I}=\{ r^{I}_1,r^{I}_2,...,r^{I}_M \}$ represents the reviews of the item $I$,
the corresponding image is denoted as $G^U=\{ g^{U}_1,g^{U}_2,...,g^{U}_N \}$.
And $y_{UI}$ means that the user $U$ has interacted with the item $I$, otherwise $y_{UI} = 0$.
As such, given the observed interaction sequence $S^U$ and user-generated content $(R^{U}, R^{I}, G^{U})$, the goal of TruthSR is to efficiently infer the probability of unobserved interactions and 
robustly provide the prediction uncertainty.
\begin{figure}[t]
\centering
\includegraphics[width=1\textwidth]{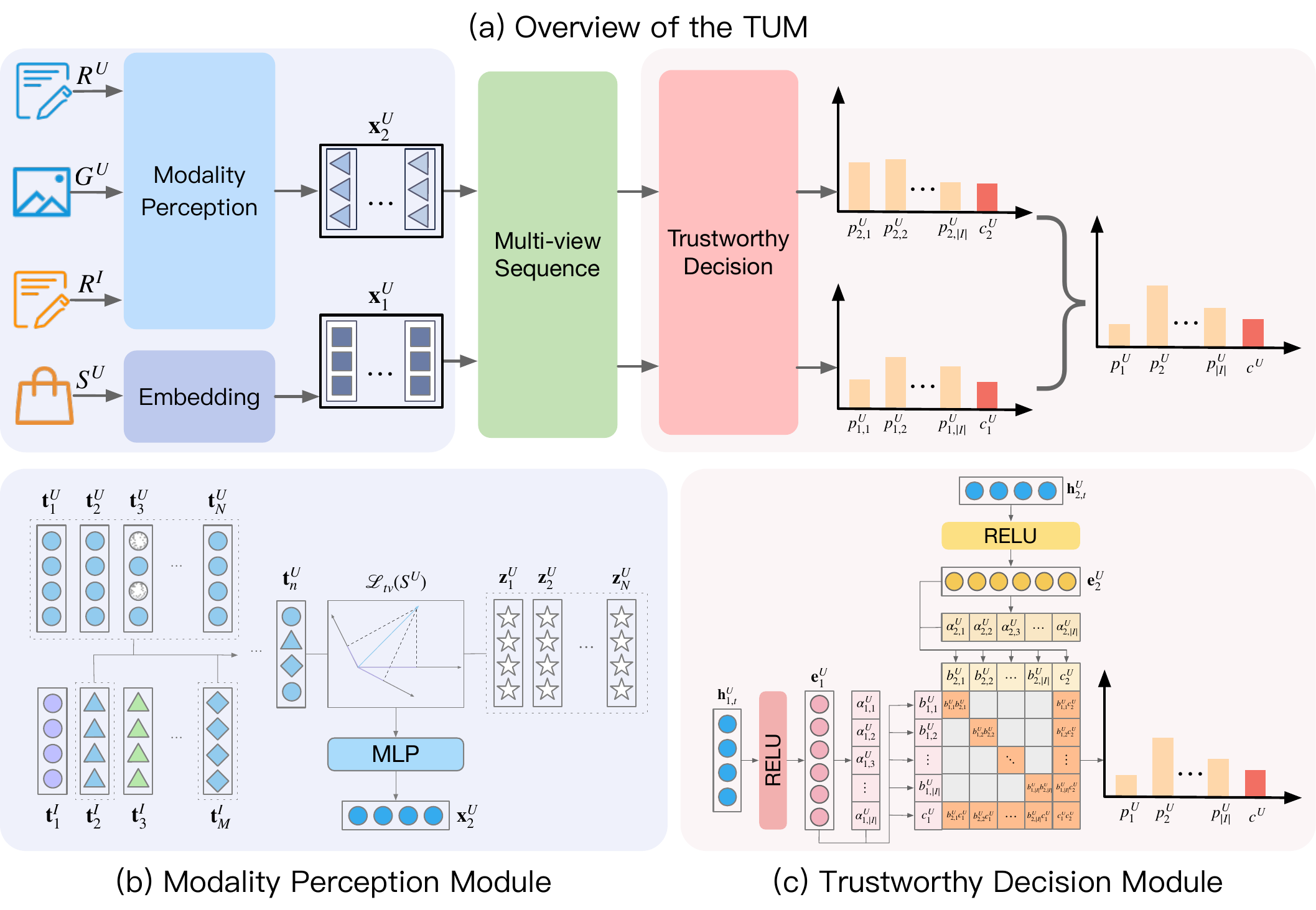}
\caption{Illustration of TruthSR. (a) shows that the model consists of embedding module, modality perception module, multi-view sequence module and trustworthy decision module. (b) shows noise reduction by capturing the consistency of the user multi-modal preferences. (c) shows the generation of trusted recommendation results and corresponding confidence levels.}
\label{fig2}
\end{figure}
\section{Model}

The overall architecture is shown in Fig.~\ref{fig2}. Its key components include the embedding module, modality perception module, multi-view sequence module and trustworthy decision module. 
Based on the observations that 1) UGC is an important complement that directly reveals the user preferences, we assume that the textual preference of $U$ over $I$ depends on the degree of similarity between $R^U$ and $R^I$, which also helps to reduce the effect of possible noise in $R^U$, while we align texts and images as much as possible to capture the user multi-modal preferences.
2) We recognize the recommendation performance from both item and user perspectives and expect to obtain consistent and trustworthy recommendation results through trusted decision networks.
\subsection{Embedding Module}
We preserve as much semantic information of UGC as possible during the embedding process. For text, we split the input text sentences into multiple vocabulary units called "token", and add special tokens such as $\texttt{[CLS]}$ at the beginning and $\texttt{[SEP]}$ at the end of the sentences. Finally, the pre-trained model outputs the hidden state of $\texttt{[CLS]}$, which represents the whole sentence for the downstream task.
Intuitively, we treat $r^U_n$ and $r^I_m$ as token sequences, i.e., $r^U_n=\{ \text{token}_t \}_{t=1}^{|r^U_n|}$, $r^I_m=\{ \text{token}_t \}_{t=1}^{|r^I_m|}$, and embed each review as follows: 
$$
    \mathbf{t}^U_n = F_t(r^U_n), \quad
    \mathbf{t}^I_m = F_t(r^I_m), \quad
    \mathbf{t}^U_n,\mathbf{t}^I_m \in \mathbb{R}^{d_t}, 
$$
where $F_t(\cdot)$ is the Bert model~\cite{ref_model1}. We obtain textual embedding $\mathbf{t}^U = [\mathbf{t}^U_1,...,\mathbf{t}^U_N] $ and 
$ \mathbf{t}^I = [\mathbf{t}^I_1,...,\mathbf{t}^I_M]$.
Accordingly, we extract visual embedding $\mathbf{z}^U = [\mathbf{z}^U_1,...,\mathbf{z}^U_N]$. For each image, we calculate the representation as follows: 
$$
    \mathbf{z}^U_n = F_v(g^U_n), \quad
    \mathbf{z}^U_n \in \mathbb{R}^{d_v}, 
$$
where $F_v(\cdot)$ is the image encoder of the CLIP model~\cite{ref_model2}.
Note that the module supports flexible embedding dimensions and multiple modalities. The pre-trained model can be regarded as an encoder that performs feature extraction on the modality.

Recommendation methods that utilize unique identifier (ID) to represent distinct items have dominated the field for more than a decade. We acknowledge its unique value and aim to retain as much information as possible about the original item ID, uninterrupted by other factors, so we consider it as an independent view. The interaction sequence $S^U$ is fed into the item embedding layer to obtain an embedding of the item's perspective:
$$ 
    \mathbf{x}^U_1 = F_i(S^U),\quad
    \mathbf{x}^U_1 \in \mathbb{R}^{d_i \times N}.
$$

\subsection{Modality Perception Module}
Users reviews of an item often only cover their limited personal preferences and do not fully reflect the overall features for the item. In addition, these reviews may contain confounding factors such as misinterpretation and advertising, the authenticity and accuracy of the reviews are questionable. By aggregating and analysing the different reviews of the same item from different users, we are able to obtain more comprehensive and diverse information about the item's characteristics. Based on the above considerations, we can re-model the user textual preferences.

Specifically, 
we extract the matching patterns between $R^U$ and $R^I$ by the co-attention mechanism~\cite{ref_model6}, which has been shown to be able to capture matching patterns between text pairs. It has three steps: first, the affinity matrix is computed as follows:
$$  
    \mathbf{A}^{UI} = \tanh ((\mathbf{t}^I)^{\top} \mathbf{M}^R \mathbf{t}^U ),\quad
    \mathbf{A}^{UI} \in(-1,1)^{M \times N},
$$
where $\mathbf{M}^R \in \mathbb{R}^{d_t\times d_t}$ is a parameter matrix. The $(i,j)$-th element of $\mathbf{A}^{UI}$ reflects the similarity between the $i$-th review of $R^I$ and the $j$-th review of $R^U$. Second, we use row-wise maximization on $\mathbf{A}^{UI}$, then use the  softmax function to generate relevance vector,
$$
    \mathbf{a}^{UI} = \text{softmax} (\text{RowMax}( \mathbf{A}^{UI} )), \quad
    \mathbf{a}^{UI} \in (0,1)^M.
$$

Last, the final relevance embedding of $R^U_n$ are calculated by performing attention aggregation on review matrices, then splice with the original text embedding:
$$
    \tilde{\mathbf{t}}^{U}_n =[\mathbf{t}^{U}_n \oplus \mathbf{a}^{UI} \mathbf{t}^I ],\quad
    \tilde{\mathbf{t}}^{U}_n \in \mathbb{R}^{d_t},
$$
where $\oplus$ represents the connection operation. 
Re-modelling each $r^{U}_n$ in $S^U$ to obtain the reconstructed user text preferences: $\tilde{\mathbf{t}}^{U} = [\tilde{\mathbf{t}}^{U}_1,...,\tilde{\mathbf{t}}^{U}_N]$.

It can be challenging to fully understand the user preferences through text alone. Images offer an intuitive visual display that can capture the user preferences for item appearance, style, colour, etc., and this information is often difficult to convey accurately through text. 
The combination of textual and visual information can enhance the understanding of the user multi-level and multi-dimensional preferences. Therefore, we integrate text and images as a new multi-modal perspective:
$$
    \mathbf{x}^U_2 = W_m[\tilde{\mathbf{t}}^{U} \oplus \mathbf{z}^U ],\quad
    \mathbf{x}^U_2 \in \mathbf{R}^{d_m \times N},
$$
where $ W_m \in \mathbb{R}^{d_m\times(d_t+d_v)}$ is the weight matrix. 
Our goal is to align the user textual and visual preferences as much as possible, so we use cross-modal loss.~\cite{ref_model7}.
The text-image pair is constructed as $ \{(r^U_i,g^U_j),o_{i,j} \}_{i,j=1}^N $, where $o_{i,j}=1$ means a matching text-image pair.
Therefore, the probability of image-text matching is defined as follows:
$$ 
    p_{ij} = \frac{\exp ((\tilde{\mathbf{t}}^U_i)^{\top} \bar{\mathbf{z}}^U_j)}{\sum\limits^N_{k=1} \exp ((\tilde{\mathbf{t}}^U_i)^{\top} \bar{\mathbf{z}}^U_k)} ,
$$
where $\bar{\mathbf{z}}^U_j = \frac{\mathbf{z}^U_j)}{||\mathbf{z}^U_j)||}$ denotes the normalised image representation, $(\tilde{\mathbf{t}}^U_i)^{\top} \bar{\mathbf{z}}^U_j$ represents the scalar projection of a text preference $\tilde{\mathbf{t}}^U_i$ onto a image preference $\mathbf{z}^U_j$.
The text-to-image loss function can be obtained by calculating the KL (Kullback-Leibler) scatter between the projection probability $p_{ij}$ and the true match probability $q_{ij}$,
$$ \mathcal{L}_{tv}(S^U)= \frac{1}{N} \sum\limits^N_{i=1} \sum\limits^N_{j=1}
   p_{ij} \log \frac{p_{ij}}{ q_{ij} +\varepsilon},$$
where $\varepsilon$ denotes a very small value to avoid numerical problems,
and the $q_{ij} = \frac{o_{i,j}}{\sum_{k=1}^N o_{i,k}}$ represents the normalised true matching probability of $(r^U_i,g^U_j)$. The loss function requires the relevant image of the text to precede the irrelevant image.

\subsection{Multi-view Sequence Module}
Existing studies show that both item IDs and multi-modal features can provide effective recommendation~\cite{ref_model3}. However, merely splicing or fusing the two together means losing valuable semantic and structural information.
So we consider the final recommendations from two complementary but independent perspectives $ \{\mathbf{x}^U_v \}_1^{|V|}$, where $v=1$ for item ID view, $v=2$ for multi-modal information view. 
Note that we are only using two views here, but this can be extended. 
Then we apply the GRU~\cite{ref_relate7} to capture sequential information of respective views:
$$
    \mathbf{h}^U_{v,t} = \text{GRU}(\mathbf{h}^U_{v,t-1},x^U_{v,t}),
$$
where $\mathbf{h}^U_{v,t}$ represents the hidden states in this view $v$. 
We use the hidden states as sequential preferences for each view, and it is well known that directly feeding hidden states into Softmax yields recommended results, but high confidence values are also often obtained for incorrectly predicted results.
We expect that consistent and trustworthy recommendation decisions can be obtained from these two perspectives.

\subsection{Trustworthy Decision Module}
We improve prediction accuracy by evaluating each view, rather than simply integrating information from the views into a single representation. Our approach dynamically assesses the credibility of the views at the level of evidence, ultimately obtaining the recommended probability for each item and the overall uncertainty of the current prediction.

The neural network can capture the evidence from input to induce classification opinions~\cite{ref_model5}, the conventional classifier can be naturally transformed into the evidence-based classifier with minor changes. We believe that recommendation is essentially a multi-classification problem with $|I|$ classes. To ensure that the network outputs non-negative values, we replace the softmax layer of the traditional classifier with an activation function layer (i.e. RELU). For the $v$-th view, these non-negative values are considered as evidence vectors $\mathbf{e}_v = [e_{v,1},...,e_{v,|I|}]$.

Our theory is based on subjective logic~\cite{ref_model4}, which defines a theoretical framework for obtaining the probabilities (belief mass) of different classes and overall uncertainty (uncertainty mass) based on the evidence, the evidence is closely related to the concentration parameters of Dirichlet distribution. 
Specifically, the parameter $\alpha _{v,i} $ of the Dirichlet distribution is induced from $e_{v,i}$ : $ \alpha _{v,i} = e_{v,i} +1$.
Then, subjective logic tries to assign a belief mass $b_{v,i}$ to each item and an overall uncertainty mass $c_v$ to the whole framework. In view $v$, the $|I|+1$ mass values are non-negative and sum to one:
$$
    \sum\limits_{i=1}^{|I|} b_{v,i} + c_v= 1 ,
$$
where $ b_{v,i} \ge 0$ and $c_v \ge 0$ represent and the probability of the $i$-th item and the overall uncertainty respectively, which are calculated as follows:
$$  
    b_{v,i} = \frac{ e_{v,i} }{D_v} = \frac{\alpha _{v,i} -1}{D_v}
    \quad and \quad
    c_v = \frac{|I|} {D_v},
$$
where Dirichlet strength is $ D_v = \sum_{i=1}^{|I|} ( e_{v,i} +1) = \sum_{i=1}^{|I|} \alpha _{v,i}$.
The belief assignment can be considered as a subjective opinion. In other words, the more evidence for the item $i$, the greater the probability that item $i$ will be assigned.

We propose an adaptive fusion method rather than assigning fixed weights to each perspective, as the quality of each perspective often varies due to different samples.
The Dempster–Shafer theory allows evidence from different sources to be combined. In our model, it combines the sets of probabilistic mass assignments for the user view and item view to obtain the joint mass, and indeed this can be extended to multiple views.
Specifically, $\mathcal{M}_1$ from the view of item ID and $\mathcal{M}_2$ from the view of multi-modal are computed together to obtain joint mass  $\mathcal{M} = \{ \{ b_{i} \}_{i=1}^{|{I}|}, c \}$:

$$
    \mathcal{M} = \mathcal{M}_1 \oplus \mathcal{M}_2.
$$
The more specific calculation rules can be formulated as follows:
$$   
    b_i = \frac{1}{1-\beta}(b_{1,i} b_{2,i} + b_{1,i}c_2 + b_{2,i}c_1)
    \quad and \quad
    c = \frac{1}{1-\beta} c_1 c_2,
$$
where $ \beta = \sum_{i \neq i} b_{1,i} b_{2,j} $ measures the conflict between the two sets, and $\frac{1}{1-\beta}$ is used for normalization.
{\color{black} According to the above,} 
the corresponding joint evidence from multiple views and the parameters of the Dirichlet distribution are induced as:
$$
    D = \frac{|{I}|}{c}, \quad
    e_i = b_i \times D \quad and \quad
    \alpha_i = e_i+1.
$$

We can obtain the estimated joint evidence $\mathbf{e}$ and the corresponding parameters of joint Dirichlet distribution $\bm{\alpha}$ to produce the final probability of each item and the overall uncertainty. We adjust the cross-entropy loss as follows:

$$ 
    \mathcal{L}(\bm{\alpha}_i)
    = \sum\limits_{j=1}^{|{I}|} y_{ij}(\psi(D_i) - \psi(\alpha_{ij})),
$$
where $\psi(\cdot)$ is the digamma function.
The loss function is the integral of the cross-entropy loss function on the simplex determined by $\bm{\alpha}_i$, which allows the correct label for each sample to generate more evidence than the other class labels. Therefore, the overall loss function for the model is:
$$
    \mathcal{L} = \mathcal{L}_{tv}(S)+
    \lambda(\sum\limits_{v=1}^{|V|}\mathcal{L}(\bm{\alpha}_{v}) + \mathcal{L}(\bm{\alpha})),
$$
where $\lambda$ is the hyper-parameter, the parameters for the Dirichlet distribution are ${\alpha}_{v}$ (for the $v$-th viewpoint) and $\alpha$ (for final viewpoints).

\section{Experiments}
To validate the experimental results, we conduct experiments on four real-world datasets. First, we present the datasets, evaluation metrics, baseline methods, and parameter configurations. Subsequently, we compare the performance of our model with that of the baseline. Finally, we conduct ablation experiments and address several questions.
\begin{itemize}
\item[$\bullet$] {\bfseries RQ1} It is unclear whether TruthSR surpasses contemporary sequential recommendation methods and multi-modal sequential recommendation methods?

\item[$\bullet$] {\bfseries RQ2} How do the different components affect the TruthSR framework?

\item[$\bullet$] {\bfseries RQ3} What is the impact of hyperparameter values on model performance?

\end{itemize}

\subsection{Experiments settings}
\subsubsection{Datasets}
We choose four real-world public datasets for our experiment, including Amazon-Beauty, Amazon-Sports, Amazon-Toys and Yelp. These datasets are constructed from Amazon review datasets~\cite{ref_dataset1}. We select product images and user reviews as additional information for all these four datasets. In reference to previous methods~\cite{ref_baseline2}, we pre-process the data and keep users and items with five or more interactions. Then, each user interactions are sorted according to the timestamp. All the interactions are regarded as implicit feedback. Statistical results of the four data sets are shown in Table~\ref{tab1}.
\vspace{-\baselineskip} 
\begin{table}
\renewcommand{\arraystretch}{1.5}
\centering
\caption{Statistics of the four datasets.}\label{tab1}
\begin{tabular}{ccccc}
\hline
Dataset&\hspace{5em} Amazon-Beauty&\quad Amazon-Sports&\quad Amazon-Toys&\quad Yelp\\
\hline
User&\hspace{5em} 22,363&\quad 35,598&\quad 19,412&\quad 30,499\\
Item&\hspace{5em} 12,101&\quad 18,357&\quad 11,924&\quad 20,068\\
Interactions&\hspace{5em} 198,502&\quad 256,308&\quad 167,597&\quad 317,182\\
Sparsity&\hspace{5em} 99.93\%&\quad 99.91\%&\quad 99.93\%&\quad 99.95\%\\
\hline
\end{tabular}
\end{table}
\vspace{-\baselineskip} 
\subsubsection{Evaluation Metrics}
To assess the performance of sequential recommendation models, we employ top-K Recall (Recall@K) and top-K Normalized Discounted Cumulative Gain (NDCG@K) with K selected from $\{10, 20\}$ two commonly used metrics. Following recommendations from~\cite{ref_evaluation1}, we evaluate model performance in a full ranking manner for a fair comparison. The ranking results are obtained over the entire item set rather than sampled ones.
\vspace{-\baselineskip} 
\subsubsection{Baselines}
We select two types of methods for comparison: robust foundational sequential recommendation methods (i.e. GRU4Rec, SASRec, Caser, BERT4Rec) and novel approaches that incorporates different sides information. It includes models utilizing attribute details (i.e. DIF-SF), review information (i.e. RNS), and multi-modal features (i.e. MMSRec). All of these models only consider item-related side information.
\begin{itemize}
\item[$\bullet$] {\bfseries GRU4Rec}~\cite{ref_relate7}: A session-based recommendation model using gate recurrent unit to better capture users behaviour and improve performance.

\item[$\bullet$] {\bfseries SASRec}~\cite{ref_baseline2}: A sequential recommendation model, which leverages self-attention mechanism to analyze entire user sequences for next-item recommendation.

\item[$\bullet$] {\bfseries Caser}~\cite{ref_relate5}: A model for personalized top-N sequential recommendation that employs convolution filters to capture both point-level and union-level sequential patterns.

\item[$\bullet$] {\bfseries BERT4Rec}~\cite{ref_baseline4}: A sequential recommendation with bidirectional encoder, which uses the Cloze task to train the bidirectional model and predicts the masked items.

\item[$\bullet$] {\bfseries DIF-SR}~\cite{ref_baseline5}: An attention-based model that moves the side information from the input to the attention layer and and decouples the attention calculation of various side information and item representation. 

\item[$\bullet$] {\bfseries RNS}~\cite{ref_baseline6}:  A novel review-driven neural sequential recommendation model by considering users intrinsic preferences and sequential patterns.

\item[$\bullet$] {\bfseries MMSRec}~\cite{ref_baseline7}: A self-supervised multi-modal method that integrates features from visual and text modalities, it employs a dual-tower retrieval architecture and self-supervised pre-training to enhance sequential recommendation performance.
\end{itemize}
\vspace{-\baselineskip} 
\subsubsection{Parameter Settings}
For all these models, we use open source code and real-world public datasets. For common hyperparameters in all models are either followed the suggestion from the methods' authors or tuned on the validation sets. We report the results of each baseline under its optimal hyper-parameter settings. We implement TruthSR with pytorch. We train them with Adagrad~\cite{ref_parameter1} optimizer for 100 epochs, with a batch size of 32 and a learning rate of 1e-2. For other hyperparameters, we iterate through all the results to find the optimal result, including: hidden\underline{\space}size $\in \{100, 200, 500, 700, 1000\}$, num\underline{\space}layers $\in  \{1, 2, 3\}$, image\underline{\space}embedding\underline{\space}dim $\in \{256, 512, 1024\}$, text\underline{\space}embedding\underline{\space}dim $\in \{256, 512, 1024\}$, results may vary for different datasets.
\vspace{-\baselineskip} 
\begin{table}[t]
\renewcommand{\arraystretch}{1.5}
\setlength{\tabcolsep}{10pt}
\centering
\caption{Overall performance comparison on the three datasets.}\label{tab2}
\resizebox{\textwidth}{!}{
\begin{tabular}{c|c|c c|c c|c c|c c}
\hline
\multirow{2}{*}{Metric}&\multirow{2}{*}{Method}&\multicolumn{2}{|c|}{Amazon-Beauty}&\multicolumn{2}{|c|}{Amazon-Sports}&\multicolumn{2}{|c|}{Amazon-Toys}&\multicolumn{2}{|c}{Yelp}\\
\cline{3-10} 
&&Recall&NDCG&Recall&NDCG&Recall&NDCG&Recall&NDCG\\
\hline
\multirow{8}{*}{k=10}
&GRU4Rec&0.0529&0.0266&0.0312&0.0157&0.0370&0.0184&0.0361&0.0184\\
&Caser&0.0474&0.0239&0.0227&0.0118&0.0361&0.0186&0.0380&0.0197\\
&BERT4Rec&0.0529&0.0237&0.0295&0.0130&0.0533&0.0234&0.0524&0.0327\\
&SASRec&0.0828&0.0371&0.0526&0.0233&0.0831&0.0375&0.0650&0.0401\\
&RNS&0.0896&0.0404&0.0532&0.0235&0.0936&0.0425&0.0676&0.0412\\
&DIF-SR&0.0908&0.0446&0.0556&0.0264&0.1013&0.0504&0.0698&0.0419\\
&MMSRec&0.0949&0.0476&0.0635&0.0323&\underline{0.1154}&0.0614&0.0749&0.0452\\
&\textbf{ours}&\underline{0.1002}&\underline{0.0493}&\underline{0.0641}&\underline{0.0342}&0.1098&\underline{0.0621}&\underline{0.0799}&\underline{0.0479}\\
\hline
\multirow{8}{*}{k=20}
&GRU4Rec&0.0839&0.0344&0.0482&0.0201&0.0588&0.0239&0.0592&0.0243\\
&Caser&0.0731&0.0304&0.0364&0.0153&0.0566&0.0238&0.0608&0.0255\\
&BERT4Rec&0.0815&0.0309&0.0465&0.0173&0.0787&0.0297&0.0756&0.0385\\
&SASRec&0.1197&0.0464&0.0773&0.0295&0.1168&0.0460&0.0928&0.0471\\
&RNS&0.1231&0.0478&0.0774&0.0299&0.1231&0.0512&0.0954&0.0472\\
&DIF-SF&0.1284&0.0541&0.0798&0.0325&0.1382&0.0597&0.1003&0.0496\\
&MMSRec&\underline{0.1341}&0.0577&0.0957&0.0423&0.1003&0.0704&0.1107&0.0525\\
&\textbf{ours}&0.1331&\underline{0.0591}&\underline{0.1021}&\underline{0.0478}&\underline{0.1079}&\underline{0.0721}&\underline{0.1147}&\underline{0.0579}\\
\hline
\end{tabular}
}
\end{table}
\begin{table}[t]
\scriptsize
\renewcommand{\arraystretch}{1.5}
\setlength{\tabcolsep}{10pt}
\caption{Results of different side-information.}\label{tab3}
\resizebox{\textwidth}{!}{
\begin{tabular}{l|c|c|c|c}
\hline
Metric&Recall@10&NDCG@10&Recall@10&NDCG@10\\
\hline
datasets&\multicolumn{2}{c|}{Amazon-Beauty}&\multicolumn{2}{c}{Amazon-Sports}\\
\hline
id&0.0534&0.0256&0.0342&0.0169\\
id+image&0.0856&0.0384&0.0522&0.0245\\
id+text&0.0899&0.0389&0.0567&0.0267\\
id+text+image&0.1001&0.0493&0.0640&0.0343\\
\hline
\end{tabular}
}
\end{table}
\subsection{Overall Performance (RQ1)}
The results of the experiment in the four datasets are shown through the Table~\ref{tab2}. Based on these results, we can see TruthSR model achieves very good results in multi-modal sequential recommendation. Furthermore, in most cases, recommendation methods that incorporate side-information(i.e. MMSRec, RNS, DIF-SF) are superior to traditional recommendation methods(i.e. GRU4Rec, Caser, BERT4Rec, SASRec). These results prove that it is feasible to use side information to improve recommendation performance. In addition, MMSRec outperforms other methods (i.e. RNS, DIF-SF) in the recommendation model that introduces side-information, indicating that the multi-modal recommendation method is better than the recommendation method that only uses text or image, but it is still weaker than our model, because our model introduces the UGC information and obtains both image and text features, making them more informative. Our proposed TruthSR model performs best in all tasks.
\begin{table}[h]
\centering
\renewcommand{\arraystretch}{1.5}
\setlength{\tabcolsep}{6pt}
\caption{Results of different components.}\label{tab4}
\resizebox{\textwidth}{!}{
\begin{tabular}{c|cc|cc|cc}
\hline
Settings&\multicolumn{2}{c|}{Beauty}&\multicolumn{2}{c|}{Toys}&\multicolumn{2}{c}{yelp}\\
Metric&Recall@10&NDCG@10&Recall@10&NDCG@10&Recall@10&NDCG@10\\
\hline
TruthSR w/o trust&0.0834&0.0384&0.0835&0.0527&0.0689&0.0356\\
TruthSR w/o filter&0.0956&0.0456&0.0978&0.0558&0.0734&0.0421\\
TruthSR w/o filter \& trust&0.1002&0.0493&0.1098&0.0621&0.0799&0.0479\\
\hline

\end{tabular}
}
\end{table}
\subsection{Ablation Study (RQ2)}
\subsubsection{Ablation Study of Side Information}
To determine the impact of various types of auxiliary information on TruthSR, we perform ablation experiments with different side information. It is widely acknowledged that considerable recommendation can be achieved using item IDs alone, we have also taken into account user reviews as text information and item images as visual information, which are common on recommendation platforms.
We design four types of experiments, as shown in the Table~\ref{tab3}. Experimental results show that using only text or image can improve the recommendation performance, proving the effectiveness of side information. In addition, the combination of two modal information can further improve the performance, which shows that our proposed fusion scheme can effectively utilize useful information from different modalities. 
\vspace{-\baselineskip} 
\subsubsection{Ablation Study of Components}
In order to investigate the impact of each component of the model on the experimental results, we design the corresponding ablation studies and conduct the experiments on the beauty, toy and Yelp datasets, respectively, and the results of the experiments are displayed in Table~\ref{tab4}.
\begin{itemize}
\item[$\bullet$] (TruthSR w/o filter \& trust) TruthSR with both filter component and trust component.
\item[$\bullet$] (TruthSR w/o filter) TruthSR with only filter component and trust component replaced by sotfmax.
\item[$\bullet$] (TruthSR w/o trust) TruthSR with only trust component.
\end{itemize}

Based on the experimental results, we can draw the following conclusions: firstly, we can see that the introduction of the two components of filter \& trust can significantly improve the performance of the model, which shows that the model we designed is effective. Secondly, the introduction of the trust component also contributes to the model, indicating that we have effectively accessed the features of the different modalities to provide more robust recommendations. Finally, the introduction of the filter improves the results considerably, which suggests that it can filter out interference in the UGC and help to improve the performance of the recommendation.

\subsection{Impact of Hyper-parameter Settings  (RQ3)}
\begin{figure}[t]
	\centering
	\begin{subfigure}{0.475\linewidth}
		\centering
		\includegraphics[width=1\linewidth]{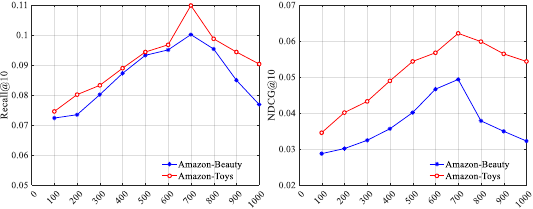}
		\caption{impact of hidden\underline{\space}size}
		\label{fig3_a}
	\end{subfigure}
	\centering
	\begin{subfigure}{0.475\linewidth}
		\centering
		\includegraphics[width=1\linewidth]{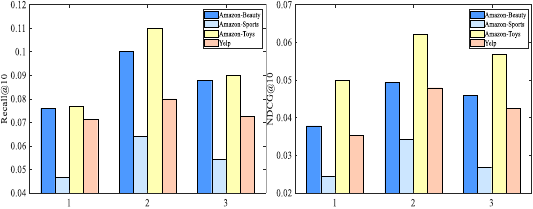}
		\caption{impact of num\underline{\space}layer}
		\label{fig3_b}
	\end{subfigure}
	\caption{impact of Hyper-parameter}
	\label{fig3}
\end{figure}

We mainly investigate the effect of two hyperparameters on the experimental results, the hidden layer dimension of the network and the number of layers. Fig.~\ref{fig3_a} presents the Recall@10 and NDCG@10 scores of TruthSR with hidden size. The results on the four datasets show that the best results are obtained when the dimension of the hidden layer is 700. Fig.~\ref{fig3_b} shows the effect of the number of network layers on the four datasets, with the best results when the number of layers is 2.

\section{Conclusion}
In this article, we explored a sequential recommendation problem that introduces user-generated content for trustworthy recommendations. We proposed a TruthSR method for this problem. TruthSR captured the consistency of user-generated content to mitigate the noise interference, it also integrated subjective user perspective and objective item perspective to dynamically evaluate the uncertainty of prediction results. Trustworthy recommendation results could be confidently applied in a variety of scenarios. The experimental results on four publicly available datasets indicated that our model outperforms the majority of existing sequential recommendation models in terms of performance. This validated the effectiveness of each component within the recommendation model. Moreover, the forward-looking research proposal should focus on explaining the model and gaining a better understanding of its underlying mechanisms, in order to provide more reliable and tailored recommendation services that meet the real needs of users. 
\subsubsection{Acknowledgment}
This research was supported by the National Natural Science Foundation of China (Grant Nos. 62133012, 61936006, 62103314, 62073255, 62303366), the Key Research and Development Program of Shanxi (Program No. 2020ZDLGY04-07), Innovation Capability Support Program of Shanxi (Program No. 2021TD-05) and Natural Science Basic Research Program of Shaanxi under Grant No.2023-JC-QN-0648.
%
%
%
%

\end{document}